\documentclass[aps,prl,twocolumn]{revtex4}
\usepackage{graphicx}
 
\begin{document}
 
\title{Comparison of Dynamical Heterogeneity in Hard-Sphere and
Attractive Glass Formers}
 
\author{David R. Reichman}
\affiliation{Department of Chemistry, Columbia University,
3000 Broadway, New York, NY 10027}

\author{Eran Rabani}
\affiliation{School of Chemistry, The Sackler Faculty
of Exact Sciences, Tel Aviv University, Tel Aviv
69978, Israel}

\author{Phillip L. Geissler} \affiliation{Department of Chemistry,
University of California, and Physical Biosciences and Materials
Sciences Divisions, Lawrence Berkeley National Laboratory, Berkeley, CA
94720}
 
\date{\today}
  
\begin{abstract}
Using molecular dynamics simulations, we have determined that the
nature of dynamical heterogeneity in jammed liquids is very sensitive
to short-ranged attractions.  Weakly attractive systems differ little
from dense hard-sphere and Lennard-Jones fluids: Particle motion is
punctuated and tends to proceed in steps of roughly a single particle
diameter.  Both of these basic features change in the presence of
appreciable attractions.  Transient periods of particle mobility and
immobility cannot be discerned at intermediate attraction strength,
for which structural relaxation is greatly enhanced.  Strong
attractions, known to dramatically inhibit relaxation, restore
bimodality of particle motion.  But in this regime, transiently mobile
particles move in steps that are significantly more biased toward
large displacements than in the case of weak attractions.  This
modified feature of dynamic heterogeneity, which cannot be captured by
conventional mode coupling theory, verifies recent predictions from a
model of spatially correlated facilitating defects.
\end{abstract}

\maketitle

Dynamical heterogeneity is perhaps the most revealing feature of
relaxation in deeply supercooled liquids.  A high-temperature fluid is
dynamically homogeneous in the sense that the local environment
restricting fluctuations of any given particle is, for all important
purposes, identical to that surrounding any other particle, even on
the short time scales of basic microscopic motions.  The distribution
$P(\delta {\bf r},t)$ of particle displacements $\delta {\bf r}$ as a
function of time $t$ provides a quantitative measure of such
uniformity.  Results of molecular dynamics simulations indicate that
$P(\delta {\bf r},t)$ is Gaussian over a wide range of displacements,
as would be expected from a mean-field perspective, for typical dense
fluids\cite{BZ}.  The microscopic environments constraining particle
motion in a glassy material are by contrast profoundly nonuniform,
even on the long time scales of large-wavelength
relaxation\cite{glotz1,glotz2,harro,onuki}.  This fact has been
clearly demonstrated by experiments that focus on dynamics of single
probe molecules\cite{vandenbout} or subsets of molecules in a pure
liquid that relax more slowly than the average\cite{israeloff,ediger}.
As a result, $P(\delta {\bf r},t)$ develops substantial weight in the
wings, reflected in appreciably nonzero values of the non-Gaussian
parameter $\alpha_2(t) = \frac{3}{5} \langle \delta r^4(t) \rangle /
\langle \delta r^2(t) \rangle^2 -1$.  Microscopy studies of colloidal
suspensions have confirmed this expectation\cite{weitz}.

Through extensive computer simulations, a detailed picture of
dynamical heterogeneity in simple jammed liquids (e.g., a binary
mixture of Lennard-Jones spheres) has developed\cite{donati}.  At low
temperatures, the majority of particles are confined within cages,
composed of neighboring particles, that may persist for very long
times.  Eventually, a rare, collective rearrangement frees a particle
from its cage, transiently allowing it to move rapidly over distances
comparable to a particle diameter.  Since such a displacement itself
facilitates local rearrangement, transient mobility appears to
propagate continuously and with some degree of directionality.
Schematic models of glassiness have been constructed with only these
features in mind\cite{BH,GCpnas,GCprl}.  They account for a surprising
variety of anomalous behaviors and yield unique scaling predictions
for the length and time scales characterizing
relaxation\cite{GCprl,garrwhite1,garrwhite2}.

These basic features of dynamical heterogeneity have little to do with
the identity of particles comprising a glassy material or with the
interactions between them.  They have been reported for many model
atomic liquids as well as for viscous silica, a network-forming liquid
which vitrifies with qualitatively different temperature
dependence\cite{glotz3}.  It is therefore tempting to presume that the
scenario sketched above is universal among supercooled molecular
liquids\cite{GCpnas,GBunp}.

When interactions between particles in a simple liquid are augmented
by strong, short-ranged attractions, spatially averaged dynamical
quantities change in nontrivial
ways\cite{dawson,cates1,cates2,sciortino}.  This situation has been
realized experimentally by adding linear chain molecules to
suspensions of colloidal particles, whose direct interactions are
almost purely repulsive\cite{pusey}.  By varying $\phi_{\rm p}$ one
can tune the fluid from purely repulsive to strongly attractive.  For
large colloid volume fraction, $\phi_{\rm c}$, and vanishing
attraction strength ($\phi_{\rm p}=0$), the suspension is in essence a
dense hard-sphere fluid with the basic phenomenology of simple
supercooled liquids.  As $\phi_{\rm p}$ increases, however, relaxation
accelerates significantly, so that a hard-sphere glass can be
``melted'' by adding attractions\cite{dawson}.  Beyond a certain value
of $\phi_{\rm p}$, relaxation becomes instead more sluggish as more
polymer is added, leading to re-vitrification at large $\phi_{\rm p}$.

\begin{figure}
\begin{center}
\includegraphics[width=8cm]{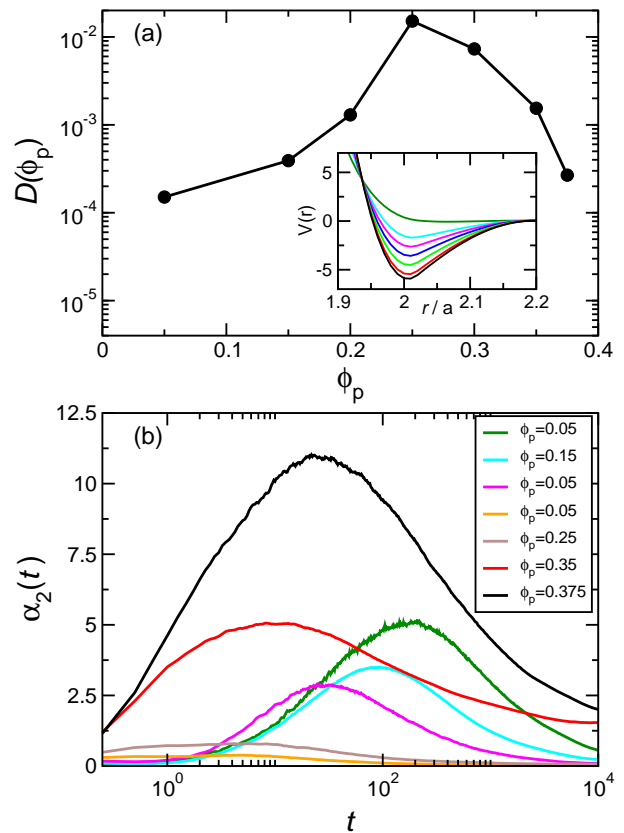}
\end{center}
\caption{(a) Diffusion constant $D$ as a function of polymer volume
fraction $\phi_{\rm p}$.  In each case, $D$ was computed from the mean
squared displacement averaged over particle size.  Inset: Potential
energy of interaction $V(r)$ as a function of the distance $r$ between
two particles with radius $a$ for the values of $\phi_{\rm p}$ we have
studied numerically.  From the least negative minimum of $V(r)$ to the
most negative minimum, these values are $\phi_{\rm p}=0.05$, 0.15,
0.2, 0.25, 0.3, 0.35, and 0.375.  (b) Non-Gaussian parameter
$\alpha_2(t)$ as a function of time $t$.  }
\label{fig:alpha2D}
\end{figure}

In this paper we examine whether the peculiar behavior of attractive
colloids reflects fundamental changes in the nature of dynamical
heterogeneity.  Although extensive computational work has been done to
characterize such heterogeneity in simple liquids, previous
simulations of materials with short-ranged attractions have not
focused on correlated microscopic motions.  
For this purpose we have adopted the model of Puertas {\em et al.} for
polymer-mediated interactions between colloids\cite{cates1,cates2}.
The effective interaction potential between a pair of colloids
separated by distance $r$, described in detail in Refs. 21-22, is
plotted in Fig.~\ref{fig:alpha2D} for the values of $\phi_{\rm p}$ we
have studied.  Interactions consist of a short-ranged repulsion
parameterized by the sum of particle radii, a short-ranged attraction
that mimics the polymer-induced depletion interaction, and a slowly
varying, long-ranged repulsion designed to prohibit phase separation
\cite{repulsion_note}.  The effective range and form of each term are
precisely as given in Refs. 21-22.  We have focused on weakly
polydisperse systems, in which these radii are drawn from a uniform
distribution, $p(R) = (2\,\delta a)^{-1} \theta[R-(a-\delta
a)]\theta[(a+\delta a)-R]$, with mean $a$ and half-width $\delta
a=a/10$. The Heaviside funcion $\theta(x)$ is defined as usual to be 1
for $x\geq 0$, and 0 for $x<0$.  In this paper all quantities with
units of length have been scaled by the mean {\em radius} $a$, and
quantities with units of time have been scaled by
$t_0=\sqrt{\frac{8a^2}{3T}}$. We have used standard methods of
molecular dynamics to propagate equilibrated systems of $1000$
periodically replicated colloidal particles, stochastically rescaling
colloid momenta every $101$ time steps to maintain a Boltzmann
distribution at temperature $T=\frac{4}{3}$.  We have investigated a
single colloid volume fraction $\phi_{\rm c}=\frac{4\pi N}{3
V}a^{3}(1+(\delta a / a)^{2})=0.55$ and several polymer volume
fractions (i.e., attraction strengths as plotted in
Fig.~\ref{fig:alpha2D}) ranging from $\phi_{\rm p}=0.05$ to $\phi_{\rm
p}=0.375$.

In order to contrast features of dynamic heterogeneity unique to
attractive systems with those generic to simple supercooled liquids,
we have selected a colloid density for which relaxation is already
very slow for $\phi_{\rm p}=0$.  The self-diffusion constant
$D(\phi_{\rm p})$, plotted in Fig.~\ref{fig:alpha2D} as a function of
$\phi_{\rm p}$, thus evinces the re-entrant behavior we have
described.  Specifically, $D$ increases by two orders of magnitude as
$\phi_{\rm p}$ approaches $0.25$ and then decreases sharply for higher
polymer concentrations.  Any meaningful comparison of relaxation
mechanisms at different $\phi_{\rm p}$ must take into account the
disparate time scales corresponding to this range of diffusivity.
Here we examine time evolution over intervals scaled such that the
overall extent of relaxation is comparable for each state.  We use several
different measures of the extent of relaxation, including $\alpha_2(t)$,
mean squared particle displacement, and the dynamic structure factor.  
Although these choices are somewhat arbitrary, they paint a consistent 
picture of changes in dynamic heterogeneity induced by short-ranged 
attractions. 

The non-Gaussian parameter, plotted in Fig.~\ref{fig:alpha2D}(b) as a
function of time for several values of $\phi_{\rm p}$, indicates that
short-ranged attractions effect changes more profound than simply a
renormalized average time or length scale\cite{cates1,cates2}.  The
peak of $\alpha_2(t)$ roughly locates the time of maximum dynamic
heterogeneity.  We denote the time corresponding to this peak as
$t^*$.  The dependence of $t^*$ on $\phi_{\rm p}$ closely mirrors that
of the diffusion constant, reflecting a global change in relaxation
time.  But the shape and scale of $\alpha_2(t)$ also change
significantly with attraction strength.  Most notably, the peak height
$\alpha_2(t^*)$ declines by nearly an order of magnitude as $\phi_{\rm
p}$ approaches 0.25, then grows rapidly for larger values of
$\phi_{\rm p}$.  This evolution strongly suggests a change in the
character of microscopic dynamics\cite{cates1}.

\begin{figure}
\begin{center}
\includegraphics[width=8cm]{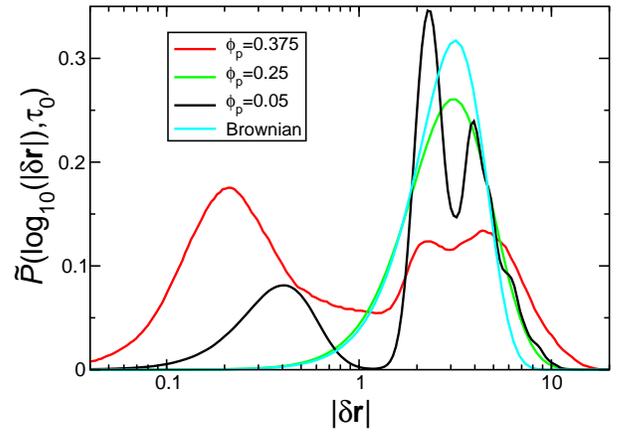}
\end{center}
\caption{ Distributions $\tilde{P}[\log_{10}|\delta {\bf r}|,\tau_0]$
of the logarithm of particle displacements $|\delta {\bf r}|$ at time
$\tau_0$.  Results are shown for polymer volume fractions $\phi_{\rm
p}=0.05$, 0.375, and 0.25, and for a Gaussian distribution of $\delta
{\bf r}$.  Times $\tau_0(\phi)$ and the width of the reference
distribution were chosen to obtain a consistent mean squared
displacement, $\langle |\delta {\bf r}|^2\rangle = 10 a^2$.  }
\label{fig:cates}
\end{figure}

The full distribution of particle displacements, or of their logarithm
($\tilde{P}[\log_{10}(|{\bf \delta r}|),t]$), over a specific time
interval (of duration $t$) provides a more detailed picture of
microscopic rearrangements\cite{cates3}.  Fig.~\ref{fig:cates} shows a
plot of $\tilde{P}[\log_{10}(|{\bf \delta r}|),\tau_{0}]$ for states
representative of weak attractions ($\phi_{\rm p} = 0.05$), strong
attractions ($\phi_{\rm p} = 0.375$), and intermediate attraction
strength ($\phi_{\rm p} = 0.25$).  Since relaxation rates vary greatly
among these three states, we have followed Cates et al.~\cite{cates3}
in comparing motion over time intervals $\tau_0(\phi_{\rm p})$
yielding the same mean squared displacement, $10 a^2$. As reflected by
$\alpha_2(t)$, the distributions at $\phi_{\rm p} = 0.05$ and
$\phi_{\rm p} = 0.375$ are highly non-Gaussian, exhibiting distinct
populations of especially mobile and especially immobile particles.
Cates et al.~\cite{cates3} have reported a similarly bimodal
distribution of $\log(|\delta {\bf r}|^2)$ for strongly attractive
colloids at much lower densities ($\phi_{\rm c} = 0.4$) and have
suggested that such bimodality is a unique feature of attractive
glasses.  Multi-peaked van Hove distribution functions, however, have
been reported several times for systems lacking short-ranged
attractions\cite{glotz4,onuki2,bagchi}.  Indeed, Fig.~2 demonstrates
that the distinction between mobile and immobile particles is in fact
{\em more} pronounced when attractions are almost negligibly weak.
Bimodally distributed particle displacements thus appear to be a
feature common to many sluggish systems.

The glassy states with weak and strong attractions also share a degree
of structure in $\tilde{P}[\log_{10}(|{\bf \delta r}|),\tau_{0}]$
within the subpopulation of mobile particles.  For $\phi_{\rm p} =
0.05$ particles clearly tend to move in discrete steps of
approximately integer multiples of a typical particle diameter, $2 a$.
This feature highlights the decoupling of diffusion and structural
relaxation in jammed liquids.  Although fluctuations in local
environment may permit a particle to move out of its cage, density
correlations persist such that the newly formed cage has a
well-defined spatial relationship with the original.  Stepwise motion
is much less pronounced at high $\phi_{\rm p}$.

By contrast, the shape of $\tilde{P}[\log_{10}|\delta {\bf
r}|,\tau_0]$ at $\phi_{\rm p}=0.25$ is nearly that of a Gaussian,
plotted for reference in Fig.~\ref{fig:cates}.  There is evidence
neither of a distinct population of immobile particles nor of stepwise
motion.  The statistics of single-particle displacements at
intermediate attraction strength thus more strongly resemble those of
a dynamically homogeneous fluid at lower density than those of a
non-attractive fluid at the same density.  The simplest conclusion is
that the $\phi_{\rm p}=0.25$ system is in essence dynamically uniform.
Evidence exists, however, that correlated motions of neighboring
particles do show signs of dynamic heterogeneity\cite{us1}.  This
situation could be expected if particle displacements were dominated
by movement of clusters transiently stabilized by
attractions\cite{us1}.  Heterogeneity associated with formation and
decay of a cluster would not have a strong signature in
$\tilde{P}[\log_{10}|\delta {\bf r}|,\tau_0]$ due to translation of
the cluster as a whole.

\begin{figure}
\begin{center}
\includegraphics[width=8cm]{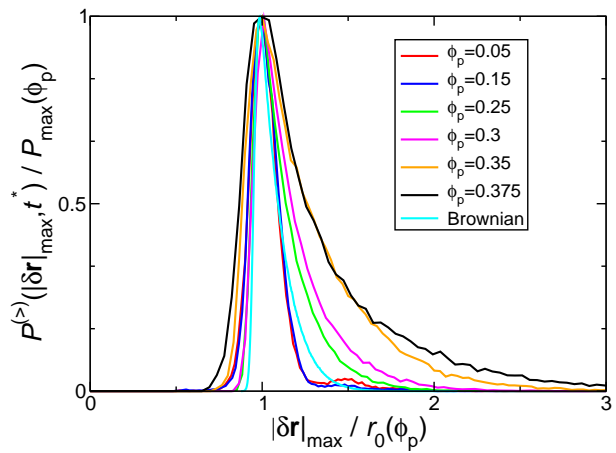}
\end{center}
\caption{ Distributions $P^{(>)}(|\delta {\bf r}|_{\rm max},t^*)$ of
maximum displacements for the 5\% most mobile particles.  We use as a
measure of mobility $|\delta {\bf r}|_{\rm max}$, the {\em largest}
displacement a particle undergoes (from its position at time $t$)
during the time interval $t$ to $t+t^*$.  Distributions and
displacements have been scaled to coincide at the peak of $P^{(>)}$.
A scaled extreme displacement distribution is also plotted for a
perfectly Brownian system of 1000 particles.  }
\label{fig:prob}
\end{figure}

The essence of a dynamic heterogeneity perspective on glassy liquids
is that relaxation over any short interval is driven by a small subset
of particles that are temporarily much more mobile than the average.
The statistics of extreme displacements should therefore be revealing
of basic relaxation mechanisms\cite{donati}.  Here we focus on
particles among the 5\% most mobile over an interval of length $t^*
\ll \tau_{0}$.  We judge mobility in this case by monitoring the
largest displacement $|\delta {\bf r}|_{\rm max}$ of a particle from
its position at the beginning of each interval.  Distributions of
maximum displacement magnitudes for these especially mobile particles,
$P^{(>)}(|\delta {\bf r}|_{\rm max},t^*)$, are plotted in
Fig.~\ref{fig:prob} for each of the polymer volume fractions we have
studied.  For purposes of comparison, we have scaled $|\delta {\bf
r}|_{\rm max}$ by its most likely value $r_0(\phi_{\rm p})$ for each
$\phi_{\rm p}$.

For reference we have included in Fig.~\ref{fig:prob} an extreme value
distribution ${\cal P}^{(>)}_{\rm B}(|\delta {\bf r}|)$ that would be
obtained for a Brownian analogue of our system.  Because maximum
displacements are not easily defined for fractal trajectories, we
consider in this case the displacement of a particle $|\delta {\bf
r}|$ from its initial position only at the end of an interval.  The
interval length is in fact arbitrary, determining only the overall
scale of displacements, which is irrelevant for the comparison in
Fig.~\ref{fig:prob}. In detail, we computed ${\cal P}^{(>)}_{\rm
B}(|\delta {\bf r}|)$ by repeatedly drawing $N=1000$ displacements
from a Gaussian distribution with unit variance in all three
dimensions, each time adding the $0.05 N$ largest values to a
histogram.  ${\cal P}^{(>)}_{\rm B}(|\delta {\bf r}|)$ is therefore a
superposition of extreme value distributions:
$$
{\cal
P}^{(>)}_{\rm B}(|\delta {\bf r}|) = 
\sum_{j=1}^{0.05 N} g_j(|\delta {\bf r}|).
$$ 
Here, $g_j(|\delta {\bf r}|)$ is the probability density for
observing $|\delta {\bf r}|$ as the $j^{\rm th}$ largest value in a
sample of size $N$.  In the limit $N\rightarrow \infty$\cite{gumbel}, 
$$
g_j(x) = {j^j \over (j-1)!}\exp{[j\overline{x}_j(x-\overline{x})-
j e^{\overline{x}(x-\overline{x}_j)}]},
$$ 
where ${\overline x}_j = \sqrt{2 \ln{(j/N)}}$.  Because convergence to
this asymptotic limit is slow, however, we have chosen to construct
${\cal P}^{(>)}_{\rm B}(|\delta {\bf r}|)$ from direct sampling.

The scaled distributions in Fig.~\ref{fig:prob} reveal a monotonic
trend toward broadly distributed mobile particle displacements as
attraction strength increases.  For weak attractions, the
large-displacement tail of $P^{(>)}(|\delta {\bf r}|_{\rm max},t^*)$
is attenuated relative to a simple Brownian fluid.  For strong
attractions, this tail is greatly enhanced.  Although the overall
shape of the full displacement distribution for intermediate
attraction strength is roughly Gaussian, statistics of the extreme
subensemble demonstrate that mobile particles nonetheless execute
larger jumps (relative to the average) than in a Brownian reference
system.  This fact is consistent with the attraction-enhanced
large-displacement tail of $\tilde{P}[\log_{10}|\delta {\bf
r}|,\tau_0]$ plotted in Fig.~\ref{fig:cates}.  

The comparison in Fig.~\ref{fig:prob} emphasizes changes in the {\em
shape} of displacement distributions.  Because we have scaled distance
by different lengths for different values of $\phi_{\rm p}$, these
results do not directly imply that transiently mobile particles
execute larger jumps in space in the case of strong attractions.  They
instead reflect the breadth of particle motion relative to the mean, a
feature that could reflect tightening of particle cages as well as
growth of facilitated regions.  More direct evidence for both of these
trends is provided by scrutinizing the trajectories of individual
particles.

\begin{figure}[t]
\begin{center}
\includegraphics[width=8cm]{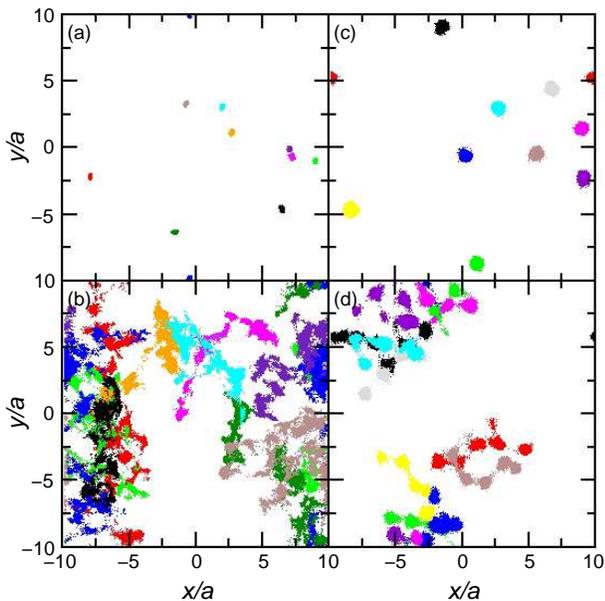}
\end{center}
\caption{ Representative trajectories of particles whose displacements
in the $xy$-plane over a time interval $\tau_\alpha$ are much smaller
or much larger in magnitude than average.  Particle traces have been
projected onto this plane for graphical simplicity.  Left panels
depict dynamics of a strongly attractive system, $\phi_{\rm p}=0.375$.
Particles among the 5\% least mobile are shown in (a), while particles
among the 5\% most mobile are shown in (b).  Right panels correspond
to a weakly attractive system, $\phi_{\rm p}=0.05$.  Again,
trajectories exhibiting extremely small displacements (c) are plotted
in the top panel.  Trajectories exhibiting extremely large
displacements (d) are plotted in the bottom panel.}
\label{fig:traj2Da}
\end{figure}

The routes traced by several particles are plotted in
Fig.~\ref{fig:traj2Da} for the smallest and largest values of
$\phi_{\rm p}$.  We have chosen the duration of displayed trajectories
to be $\tau_\alpha$, the time required for correlation of density
fluctuations over length scales comparable to a particle diameter to
decay by a factor of $e$.  Dynamics over this time scale, which is
much larger than $t^*$ in both cases, exhibit the full character of
particle motion.  This choice also allows a straightforward comparison
of dynamics for weak and strong attractions, since the corresponding
values of $\tau_\alpha$ are very similar ($\tau_\alpha\approx 8\times
10^3$ for $\phi_{\rm p}=0.05$ and $\tau_\alpha \approx 10^4$ for
$\phi_{\rm p}=0.375$).  The depicted excursions of extremely immobile
(top panels of Fig.~\ref{fig:traj2Da}) and extremely mobile (bottom
panels) particles reinforce the dynamical features we have gleaned
from probability distributions.  The 5\% least mobile particles
exhibit only small flucutations about their initial positions, both
for $\phi_{\rm p}=0.375$ (a) and for $\phi_{\rm p}=0.05$ (c), even
over this long time scale.  The cages that constrain these particles'
motion are clearly smaller in the case of strong attractions.  The 5\%
most mobile particles, on the other hand, move several particle
diameters during the same interval.  For the case of weak attractions
in panel~(d), the discrete nature of particle motion and correlations
between subsequent cages are immediately evident.  Trajectories are
qualitatively different for the case of strong attractions in
panel~(b).  Here, mobile particles explore much more diffuse regions.
Domains of facile particle motion are clustered in space and markedly
elongated, with asymmetries apparently correlated over several
particle radii.

Most of the qualitative changes in dynamical heterogeneity we have
reported can be understood as consequences of changing spatial
patterns of structural defects\cite{footnote}.  Models based on
dynamical heterogeneity typically assume that the subtle defects which
enable local relaxation are sufficiently sparse as to be statistically
independent (despite significant correlations in defect dynamics).  We
have proposed that an important effect of short-ranged attractions is
to introduce non-negligible spatial correlations among such
facilitating entities\cite{us2}.  In particular, defects should
aggregate with increasing attraction strength as an indirect result of
particle clustering.  Microscopic regions of mobility thus grow in
size, but become still more sparse if their overall concentration
(loosely analogous to free volume) is held fixed.  This picture
accounts for the broadening distributions of mobile particle
displacements we have computed.  Particle trajectories depicted in
Fig.~\ref{fig:traj2Da} make the agreement especially vivid.  The
limited spatial extent of facilitating defects in a hard sphere glass
cuts off the range of available displacements.  Clustering of these
defects as attractions are introduced provides increasingly extended
loose regions for mobile particles to explore.  Our numerical results
provide strong evidence for segregation of jammed and unjammed regions
of attractive liquids at high density.  Such segregation is an obvious
feature of attractive colloids at low $\phi_{\rm c}$, which form
stable (though likely nonequilibrium) gel-like networks.  In that
case, the spacing between dense regions of the network establishes a
minimum length scale for dynamic heterogeneity.  It is remarkable that
remnants of this behavior persist at high packing fraction, where
spatial heterogeneity of liquid structure is subtle.

In summary, the change in dynamics induced by short-ranged attractions
in a dense model liquid is dramatic even on the microscopic scale.
Our results reveal three distinct regimes of dynamical heterogeneity.
For weak attractions, mobilized particles make discrete jumps between
cage structures, which may remain correlated over many jump times.
For a range of intermediate attraction strengths, Gaussian particle
displacement statistics suggest instead very fluid and uniform motion.
The role of attractions in this regime, we suggest, is to bind small
transient clusters which move on a time scale comparable to their
lifetimes.  Strong attractions restore some discreteness of particle
displacements, presumably because transient clusters are too large to
move appreciably on pertinent time scales.  We thus conclude that
``attractive'' glassiness is driven by spatial redistribution of
facilitating defects.  Although extended loose domains permit large
particle displacements, their growth depletes mobility in surrounding
areas, which in turn inhibits relaxation of domain interfaces.  We are
pursuing further calculations to confirm this clustering of mobility
in dense environments and to characterize the correlated fluctuations
underlying fluidity at intermediate attraction strength.

We would like to thank Laura Kaufman for useful discussions.  We
acknowledge the NSF (D.R.R.), DOE (P.L.G.), and the United
States-Israel Binational Science Foundation (grant number 2002100 to
D.R.R and E.R.) for financial support.  D.R.R. is a Camille Dreyfus
Teacher-Scholar and an Alfred P. Sloan Foundation Fellow.

\bibliographystyle{prsty}

\begin{thebibliography}{10}
\bibitem{repulsion_note}
The effect of such a term on the type of heterogeneous motion discussed here
  has not been unambiguously demonstrated. Nor is it clear that such a
  repulsion exists between colloidal particles studied in experiments.
\end{thebibliography}

\end{document}